# IMPROVING ENERGY EFFICIENCY IN MANETS BY MULTI-PATH ROUTING

Hassanali Nasehi[1], Nastooh Taheri Javan[1], Amir Bagheri Aghababa[1] and Yasna Ghanbari Birgani[2]

[1]Department of Engineering, East Tehran Branch, Islamic Azad University (IAU)
Tehran, Iran
[2]Department of Industrial Engineering, Tarbiat Modares University (TMU)
Tehran, Iran
`hanasehi@yahoo.com, nastooh@aut.ac.ir, amir_baqeri_aqababa@yahoo.com,`
`yasna.ghanbari@modares.ac.ir`

## ABSTRACT

*Some multi-path routing algorithm in MANET, simultaneously send information to the destination through several directions to reduce end-to-end delay. In all these algorithms, the sent traffic through a path affects the adjacent path and unintentionally increases the delay due to the use of adjacent paths. Because, there are repetitive competitions among neighboring nodes, in order to obtain the joint channel in adjacent paths. The represented algorithm in this study tries to discover the distinct paths between source and destination nodes with using Omni directional antennas, to send information through these simultaneously. For this purpose, the number of active neighbors is counted in each direction with using a strategy. These criterions are effectively used to select routes. Proposed algorithm is based on AODV routing algorithm, and in the end it is compared with AOMDV, AODVM, and IZM-DSR algorithms which are multi-path routing algorithms based on AODV and DSR. Simulation results show that using the proposed algorithm creates a significant improvement in energy efficiency and reducing end-to-end delay.*

## 1. INTRODUCTION

MANETs are such networks in which there are no infrastructures [1]. Therefore, all network functions such as routing and data transfer are performed by the nodes themselves and by cooperating with each other.  These types of networks, the routing process has its special difficulty and complexity due to the high mobility of nodes and dynamic network topology. From the beginning of the mobile networks, many routing algorithms for these networks have been suggested. Some of those algorithms are AODV and DSR that have more popularity and acceptability than other algorithms [2]. Both of these algorithms relates to the class of on-demand routing algorithms.   In the on- demand routing algorithm, the path discovery process begins when a node has a packet to send without any valid path to its specific destinations.

Among the mass network routing algorithms in mobile ad hoc networks, multi-path algorithms have found their place [3].  In these algorithms, it tends to discover several paths between source and destination rather than finding a path between them .The main advantage of this idea is that the time consuming process is executed less times to discover paths. And when one of the routes faces with failure, it is possible to use one another discovered path quickly. Among the multi-path routing algorithms in MANET, Some of them after discovering some paths between origin and destination start sending data through several paths simultaneously, in order to reduce latency and increase end to end bandwidth.

One of the most important issues in any kind of multi-path routing algorithms is how to select multiple paths. These algorithms, try to use more discrete paths as possible to increase the reliability from one side and reduce shared resources, increase bandwidth and reduce latency from another side. In fact, these algorithms, in their best status, prefer to use node-disjoint paths. In node disjoint paths, there is no node between two joint paths, and, therefore paths are completely independent and they don't have any shared resources. Thus, paths' bandwidth doesn't have anything in common, and by the deterioration of a node or a connection, at most one of the paths will be lost.

Up to now, by discovering and choosing node-disjoint paths, everything seems to be good. However, by focusing on the essence and the structure of MANET, a new problem comes up. As we know, in MANET, there are inherent problems such as Exposed Terminal Problem and Hidden Terminal Problem. To resolve these problems, CSMA/CA protocol has been proposed [4] which is used for accessing the channels in 802.11 standard. In this protocol, due to exchanging the RTS and CTS messages between nodes, some nodes are forced to avoid sending information. This fact will increase the end-to-end delay [5].

For instance, figure 1 shows a hypothetical network in which ten nodes is exhibited. Also, the radio range of each node is specified and the dash lines indicate direct connection between two nodes. In other words, the dash line between two specific nodes means that two nodes are in radio range of each other. In this network, between nodes S and D, there are two distinct paths S-I1-I2-I3-I4-D and S-I5-I6-I7-I8-D (upper and lower paths) in which communication and data transmission through one path are not completely independent from another path. In this state, the end-to-end delay of each path depends on the traffic of the other path. This is because of the exchange of RTS and CTS messages among nodes of network for avoiding collisions and salvation of Exposed Terminal Problem and Hidden Terminal Problem. As a result, some terminals of a path should postpone their sending process in order to receive CTS from a node in the opposite path, for instance.

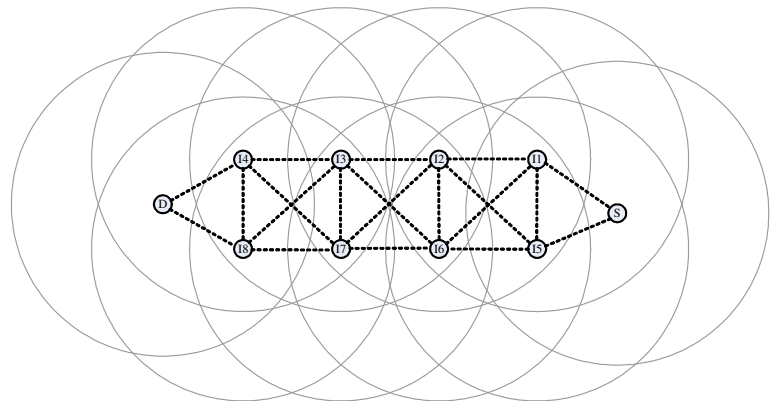

**Figure 1.** Node-Disjoint Paths.[5]

To resolve this problem, several methods have been proposed in which directional antennas are used [6]. Also we can use zone-disjoint paths instead of node-disjoint paths [5]. Two routes with no pair of neighbor nodes are called zone-disjoint in terminology.

In this paper, we have tried to resolve this problem somehow, by modifying the AODV routing algorithm and using omni-directional antennas. In our previous work [7] we focused on end-to-end delay, but in present work we have tried to improve energy efficiency in MANET.

This article contains following parts. In the second part of this article, we have analyzed earlier works around the same topic. In Section 3, we go through the details of the proposed algorithm. Also, in section 4, we explain the simulation results of our approach. Finally, in Section 5, a brief conclusion of the work is represented.

## 2. RELATED WORKS

A lot of approaches have been made in multi-path routing in MANET [8, 9, 10]. Moreover, single-path algorithms have been applied in such algorithms due to their good performance. One of these single-path algorithms is AODV. Also, AOMDV algorithm has been suggested based on AODV [11]. AOMDV algorithm has become popular because of its simplicity. This algorithm tries to discover distinct disjoint-paths between origin and destination. In this state, source creates a path request message and broadcasts it to all of its neighbors. Unlike single-path algorithm in which analogous messages are deleted, this algorithm doesn't delete repeated messages. Since the distinct paths should be discovered, each of these source neighbors will be the beginner of a new feasible path. Source neighboring nodes rebuild request package and re-broadcast it by entering their address in RREQ. Each intermediate node receives and analyzes this packet. If middle nodes have not responded to such request, they receive the packet and rebroadcast it with new number of Hop Counts. But if such a packet was previously replied, and if the following conditions are satisfied, the packet will be accepted and inserted in the node table.

1. If the packet is received a from different neighbor node
2. The packet is received from this source neighbor node which we have not had any path from it so far. (This packet is a presenter of a new path).
3. If the number of hop counts is less than the existing values along the way.

The destination node, upon receiving the path request packet, sends a desired number of path reply message to the request packets for this request. Therefore, paths can be elicited at the source point.

AODVM [12] is also based on AODV, and can find only node disjoint paths. The scheme proposed in AODVM takes advantage of reliable nodes along with multipath routing to construct a reliable path between two nodes. Packet salvaging can be used in multipath routing to provide improved fault tolerance.

Zone disjoint multi-paths are implemented with using directional antennas [6]. In this method, nodes are constantly inserting neighbor's information including power and neighbor signal angle, in a table. Among the existing paths, node disjoint paths are selected, and then zone distinction of the paths will be determined by each node's tables. This work will be constantly repeated to always ensure finding zone disjoint paths. One of the disadvantages of this method is that the directional antennas must be used. However, multi-directional antennas are used in most ad-hoc networks and directional antennas are not available .

In [13], the idea of discovering zone disjoint paths in Source Routing algorithms with using Omni directional antennas has been explored, and in [14] this idea has been implemented and applied in DSR base routing algorithm. In this algorithm, after discovering the zone disjoint paths from source to destination, it has been tried to send the information to destination via discovered paths simultaneously. Also in this algorithm, the idea of counting the number of active neighbors is performed in order to discover zone disjoint paths. In our previous work [7] we implemented this method in AODV.

## 3. SUGGESTED ALGORITHM

The suggested algorithm has been designed and implemented based on AODV algorithm. The AODV algorithm is considered to be in the class of on-demand routing algorithms in which routing process takes place hop by hop. In this way, each node has a path table in which received packet's information are saved.

As it is mentioned in the introduction, the proposed algorithm tries to discover zone disjoint paths between source and destination in order to send information simultaneously. If there is possibly no neighboring between two nodes in two distinct paths, the paths are called area distinct. Briefly, the proposed algorithm counts the number of active neighbors for each path, and finally it chooses some paths for sending information in which each node has lower number of active neighbors all together. Here, active neighbors of a node are defined as nodes that have previously received the RREQ. There is this possibility that source and destination choose another path with nodes to exchange information; thus, information exchanging depends on this path. In fact, these two nodes are on two disjoint but adjacent paths.

### 3.1. Necessary alterations in the AODV algorithm

In most of the implementations of the AODV algorithm, the middle nodes maintain a Route Cache table in which they put paths that have been discovered during the course of the path discovery. Thus, if an middle node receives a packet containing path request from a predetermined source, it will return a path reply packet to the source. However, in the proposed algorithm, middle nodes don't need to maintain Route Cache tables. Therefore, more path request packets will reach destination. In fact, all the path request packets move from source to destination.

In addition, in the suggested algorithm, each node must put the received RREQ specifications in the table which is called RREQ_Seen, in order to respond to the neighbor queries properly. Also, to count the active neighbors in each path, in the RREQ_Seen table, each node has a field named "the number of active neighbors after sending RREQ" that this field is briefly called After_A_N_C in this article. Also, the ActiveNeighborCount field name is added to the headers of RREQ and RREP to make next nodes of the path aware of the number of neighboring nodes in traversed nodes. Finally, two new packet as RREQ_Query and RREQ_Query_Reply are added to the path discovery process to perform the query process. More thoroughly, the query initiator node places current RREQ profile information into RREQ_Querypacket and sends it to its neighbors. If nodes themselves are the answer of the auery process they turn back a RREQ_Query_Reply to the initiator.

### 3.2. The suggested algorithm's procedures

When a node is about to send data to a specific destination and it does not find a valid path to its destination, the node runs the path discovery process by producing and sending RREQ packet to its neighbors. In this RREQ packet, the initial value of zero will be assigned to ActiveNeighborCount field. Therefore, source neighbor nodes receive RREQ packet, set their names as the founders of one of the paths and reversely put the path specifications into the path table. But before resending the RREQ packet, the neighbor nodes request query path from their neighbors. In fact, they ask all their neighbors: " Have you seen a RREQ with this specification?" Then they increase the value of ActiveNeighborCount in RREQ packet for those neighbors which have a positive answer to this question. For this query, nodes use some packets with titles of RREQ_Query and RREQ_Query_Reply. Actually, the query node sends the RREQ_Query packet to its neighbors and after specific time period (which is calculated by a clock) waits for neighbors' responses to the question. On the other hand, all neighboring nodes are required to search the specification of RREQ in RREQ_Seen table after receiving the query packet. If neighboring nodes have already observed this RREQ, they reply a positive response to the query node. The response to query is performed by the production and transmission of RREQ_Query_Reply packet. Finally, after the time expiration, the node that has created the query broadcasts the RREQ packet to continue the discovery process.

Once again, we analyze the behavior of queried node. Since the repeated RREQ packets aren't removed in discovery of multiple paths, it is possible for a node to receive the RREQ packet for the second time. Therefore, it initiates the query process to discover new possible neighbors for the second time to. But obviously, only new neighbors need to consider this query important and old neighbors shouldn't answer to this repeated query. Thus, those nodes that receive the query packet keep the address and details of the query node and the queried RREQ packet in Query_Seen table. If a node receives a query packet for the first time, it sends a RREQ_Query_Reply packet to inform query node after recording a query's specifications. But if this query has already been received from the same node, it is not noticed.

Now, we focus on another aspect of query node. It is possible that a node to be queried by a new node after receiving a RREQ, performing query, updating the ActiveNeighborCount field in RREQ packet, and finally sending the RREQ (Further, this scenario is described in an example.) In fact this new neighbor is not considered in computing of the node. To resolve this problem, a field as After_A_N_C is added to RREQ_Seen table of each node. Now, after rejecting RREQ packet, if the node has been queried about the RREQ before the RREP arrivals, it adds one unite to the value of after_A_N_C field for each query. Noticeably, this node is still answering the query positively so that the query node can calculate accurately. After these measurements, when the RREP is sent from the source to the destination, in the middle of the path, each node adds the value of After_A_N_C in its RREQ_Seen table to ActiveNeighborCount field in RREP packet.

Thus, when a RREP packet reaches to the source, its ActiveNeighborCount field has already counted the exact number of this path's active neighbors. At this point, source can choose those RREPs from received ones that have the lowest ActiveNeighborCount and send information simultaneously through those paths. For this purpose, the source sets a clock after receiving the first RREP and waits for receiving the rest of RREP. After the timer expires, it chooses paths with less ActiveNeighborCount .

### 3.3. The suggested algorithm's pseudo-code

For a better understanding of the proposed algorithm, the source nodes function pseudo code in figure 2, destination node function pseudo code in figure 3, and pseudo code of the middle node function are presented in Figure 4.

> 1. If you have data to send and you don't have a valid path to that destination, broadcast the RREQ packet.
> 2. Wait for RREP to arrive.
> 3. In case of receiving the first RREP, wait for a while and then choose those paths among received paths that have the lowest number of neighbors. After wards, start sending data via this path.

**Figure 2.** Source node pseudo code in the suggested algorithm

> 1. Send the corresponding path PPEP for all the nodes that you have received the RREQ packets.

**Figure 3.** Destination node pseudo code in the proposed algorithm

1. If you received the RREQ packet and this packet is acceptable, do the following steps. Otherwise, dismiss the packet.
   a. Put this packet's specification into the RREQ_Seen table.
   b. Prepare the RREQ_QUERY packet and assign it a value.
   c. There is a question on this packet that asks: Have you seen such a request packet before?
   d. Send the RREQ_Query packet to your neighbors
   e. Wait a specific period of time for your neighbors to reply
   f. Increase the ActiveNeighborCount with regard to the number of accepted replies.
   g. Rebroadcast the RREQ packet
2. When you received the RREQ_Query packet, perform the following actions:
   a. With regard to the RREQ_Seen table, if you have not seen this RREQ before, dismiss the packet and don't consider it.
   b. According to the REQ_Seen table, if you have seen this RREQ before, inform the query node by sending a RREQ_Query_Reply packet then add one unite to the After_A_N_C field of the corresponding RREQ in its RREQ_Seen table.
3. If you have received the RREQ_Qeury_Reply packet, add one unite to this RREQ's AvtiveNeighborCount field.
4. When you receive the RREP packet, add the corresponding after_a_n_c to activeneighborcount field of RREP packet and send it.

**Figure 4**. middle node pseudo code in the proposed algorithm

For a better understanding of the algorithm, consider the hypothetical network in figure 5. In this network, the line between two nodes means that these two nodes are in each other's radio range.

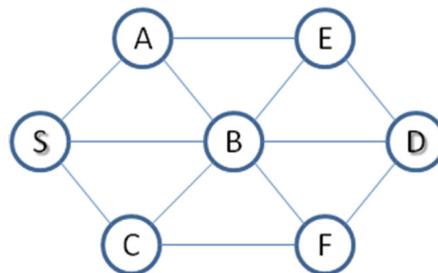

**Figure 5.** Network Topology of our example.

Suppose in this example, the node S as a source node wants to send data to node D as the destination but it does not know the path to the destination node. Therefore, node S sends the RREQ packet with the zero value of ActiveNeighborCount field to all. In the first stage, A, B and C receive the packet and insert its specifications in RREQ_Seen table along with the initial value of zero for After_A_N_C field. Then they add their address as the founder of a path in the RREQ packet. In addition, they begin the query procedure according to the proposed algorithm. For this purpose, they send the RREQ_Query packet for their neighbors and wait for their neighbors to respond by setting a clock. After making inquiries, node A and C only recognize the node B in their neighboring and add one unite to the ActiveNeighborCount field in RREQ Packet, but node B recognizes the neighboring of the node A and C. Also, it adds two units to ActiveNeighborCount field in RREQ Packet. Then all three nodes of A, B and C propagate the RREQ packet to complete the discovery process. This process is shown in figure 6.

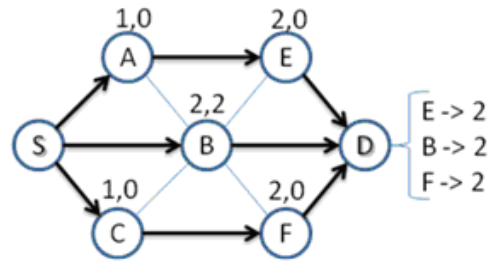

**Figure 6.** First Phase of our Algorithm.

Each node has two numbers written in its above, the number on the left shows the value of ActiveNeighborCount field and the number on the right shows the After_A_N_C field value exactly before re-broadcast RREQ in the corresponding RREQ_Seen table. In the next step, these RREQs reach other nodes. Because of the simplicity of the figures 6 and 7, we refused to show other versions of the RREQ which have been delivered to nodes by other paths. The node B delivers the first RREQ to D, and the destination node replies by sending a RREP packet to the source. Noticeably, this path's ActiveNeighborCount is equal to 2. However, RREQ reaches to the E and F through the paths in the corresponding figure. These two nodes begin the query processes separately, recognize the node B in their neighboring, and add one unite to the existing ActiveNeighborCount in RREQ. Factually, the nodes E and F send the RREQ_Query packets to node B to perform the query process. Also, because node B has propagated this RREQ before, it has a positive answer for both queries and it sends the RREQ_Query_Reply packet to each. Furthermore, node B finds two new neighbor (E and F)after this inquiry. Therefore, after sending the query reply, node B adds one unite to After_A_N_C field of current path in RREQ_Seen table per each query. This procedure is well shown in Figure 6, and we can see that the value of After_A_N_C field of node B is equal to 2. Figure 6 shows the network status at the moment that all RREQs have reached the destination. As it is indicated in the figure, all three RREQs have been received in destination with an equal amount of ActiveNeighborCount which is 2. At this stage, the destination receives the RREQ, creates its relevant RREP Packet, fills the corresponding RREQ with the same amount of the existing ActiveNeighborCount field in RREP, and then sends the RREP packets to source.

After receiving each of these RREPs, middle nodes must add the After_A_N_C value from RREQ_Seen ActiveNeighborCount table to field in RREP. This is shown in figure 7. In this figure, the corresponding sum action of each node is shown above the nodes. As it can be seen, nodes A, E, F and C in this scenario does not add any value to the ActiveNeighborCount field in RREP, but node B adds two units to ActiveNeighborCount in its own RREP. In the end, RREP packets are delivered to the source (node S).

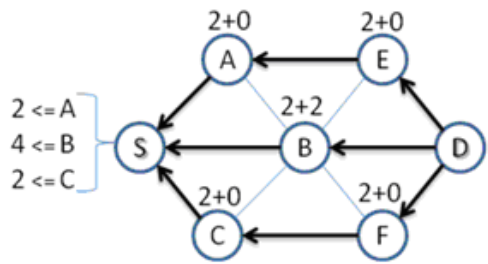

**Figure 7.** Second Phase of our Algorithm.

By receiving the first RREP and setting a timer, the source node waits for a certain period of time. After timer expiration, the source node sorts the received RREPs in an ascending order.

Then it selects the required number of RREPs from the beginning of queue and begins to send the information concurrently through the selected paths. (In fact, those paths are selected that have less ActiveNeighborCount.) For example, we suppose that the origin is determined to send data concurrently to the destination through two paths. In this case, as it is shown in figure 8, both the SAED and S-CFD paths can be selected.

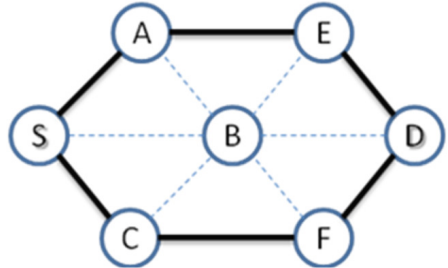

**Figure 8.** Selected Paths in our example

## 4. SIMULATION

The results of simulations and a comparison between the proposed algorithm and other existing algorithms are presented in this section. For this purpose, the following algorithms have been compared with each other in various scenarios:

- The proposed multi-path routing algorithm, which is presented as ZD-AOMDV in graphs and results.
- AOMDV [11]
- AODVM [12]
- IZM-DSR [14]

### 4.1. Simulation Environments

In this study, GLOMOSIM is used for the simulation [15]. For this purpose, we have compared the proposed algorithm with AOMDV algorithm in various scenarios. Conditions, simulation environment and simulation results are presented in this section. In these simulations, both algorithms use three paths for sending data simultaneously.

50 nodes with radio range of 250m in an environment with the dimension of 750x750m are used for simulation. In such status, nodes have random movement with using the Random Waypoint mobility model. In this model, each node randomly selects a point as a destination. After the node reaches to destination, it stays at the same point for the duration of Pause Time and again it repeats the same action. In all simulations, we consider one second for Pause Time. Nodes also use IEEE 802.11 MAC layer protocol; moreover, nodes use RADIO-ACCNOISE standard radio model for sending or receiving information. CBR model is used for traffic model. The time duration for each simulation has been considered 300 seconds and the recorded results is the outcome of an average 25 times for each simulation.

### 4.2. Simulation Metrics

Five important performance metrics were evaluated in our simulation: (i) End-to-End Delay Average –this includes all possible delays caused by buffering during route discovery phase, queuing at the interface queue, retransmission at the MAC layer, propagation and transfer delays. (ii) Packet Delivery Ratio, (iii) Routing Overhead Ratio – the number of routing control packets per each data packet. (iv) Energy Consumption. (v) Number of Dead Nodes.

### 4.3. Simulation results

*A) Packet Delivery Ratio*

In figures 9-12, the 4 algorithms have been compared with each other in terms of packet delivery rate according to the results of the simulation. In all algorithms, the packet delivery rate will decrease by increasing the maximum speed of nodes .Moreover, this decrease is because of the dynamic nature of topology and the increase of network connections termination rate.

Figure 9 shows packet delivery ratio versus max speed of nodes in random way point model, figure 10 shows packet delivery ratio versus pause time of nodes in random way point model, figure 11 shows packet delivery ratio versus number of data sources (number of connections) and figure 12 shows packet delivery ratio versus offered traffic.

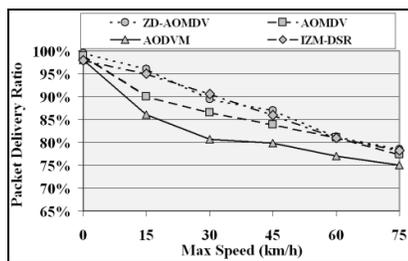
**Figure 9.** Packet delivery ratio vs. Max Speed.
(Pause time=15ms, No of Src=1, Traffic= 35Kbps)

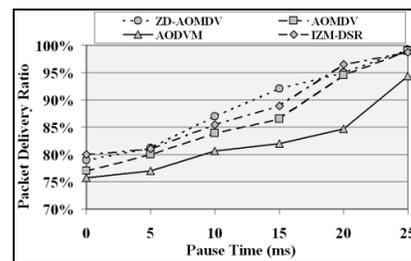
**Figure 10.** Packet delivery ratio vs. Pause time
(Max speed=40Km/h, No of Src=1, Traffic=35Kbps)

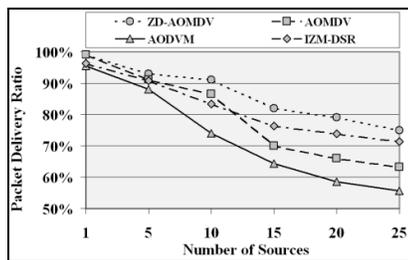
**Figure 11.** Packet delivery ratio vs. No. of Sources
(Max speed=40Km/h, Pause time=15ms,Trffic=35Kbps)

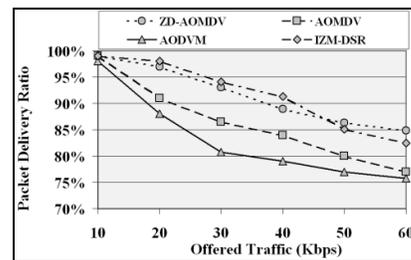
**Figure 12.** Packet delivery ratio vs. Offered Terrafic
(Max speed=40Km/s, Pause time=15ms, No of Src=1)

*B) End-to-end Delay*

In figures 13-16, the four algorithms have been compared in terms of the average of end-to-end delay. As it can be seen, with increasing the maximum speed of nodes the end-to-end delay average of packets increases correspondingly. With focusing on figure 13, the ZD-AOMDV algorithm reaches to less end-to-end delay than the other algorithms. As it has been described, the route discovery phase of the ZD-AOMDV algorithm is run with more delay. Instead, this algorithm compensate for this delay while it sends data. As a result, the end-to-end delay for algorithm ZD-AOMDV is less than other algorithms.

International Journal of Wireless & Mobile Networks (IJWMN) Vol. 5, No. 1, February 2013

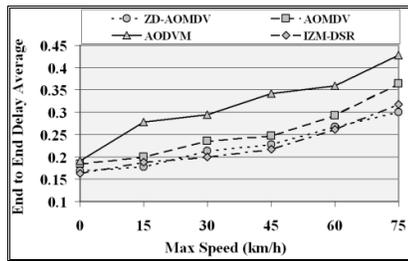

**Figure 13.** End-to-end delay vs. Max speed
(Pause time=15ms, No of Src=1, Traffic=35Kbps)

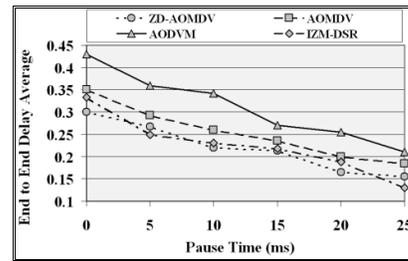

**Figure 14.** End-to-end delay vs. Pause time
(Max speed=40Km/h, No of Src=1, Traffic=35Kbps)

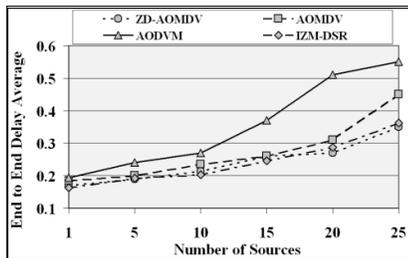

**Figure 15.** End-to-end delay vs. No. of Sources
(Max speed=40Km/h, Pause time=15ms, Traffic=35Kbps)

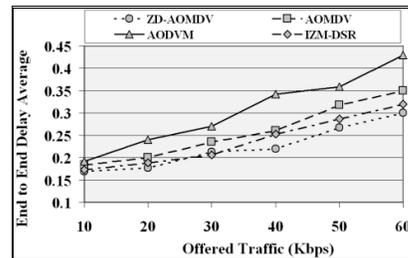

**Figure 16.** End-to-end delay vs. Offered Traffic
(Max speed=40Km/h, Pause time=15ms, No of Src=1)

*C) Routing Overhead*

Also in this section, the four algorithms have been compared in terms of routing overhead. The ZD-AOMDV and AOMDV algorithms send data through three paths simultaneously. The results of this simulation model have been displayed in figures 17-20. In this case the proposed algorithm has greater control overhead ratio than other algorithms.

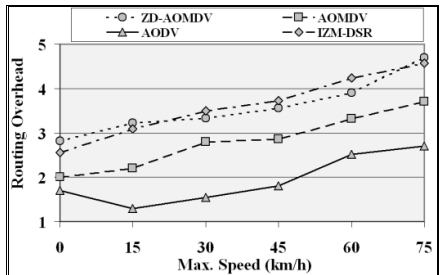

**Figure 17.** Routing overhead vs. Max speed
(Pause time=15ms, No of Src=1, Traffic=35Kbps)

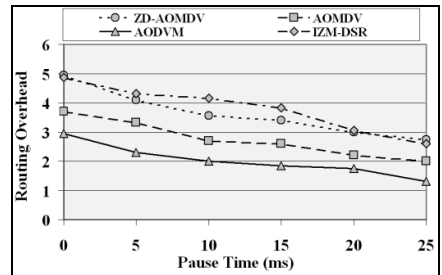

**Figure 18.** Routing overhead vs. Pause time
(Max speed=40Km/h, No of Src=1, Traffic=35Kbps)

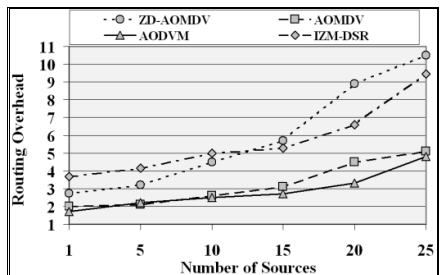

**Figure 19.** Routing overhead vs. No. of Sources
(Max speed=40Km/h, Pause time=15ms, Traffic=35Kbps)

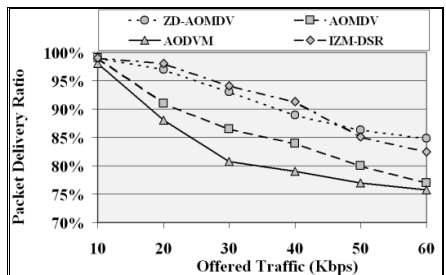

**Figure 20.** Routing overhead vs. Offered Traffic
(Max speed=40Km/h, Pause time=15ms, No of Src=1)

*D) Number of Dead Nodes*

In figure 23 the number of dead nodes in the network is depicted, and if we consider the network lifetime to be the period in which at least half of the nodes in the network are alive, then it can be realized from this figure that by using ZD-AOMDV instead of using AODVM protocol, network lifetime will nearly be doubled.

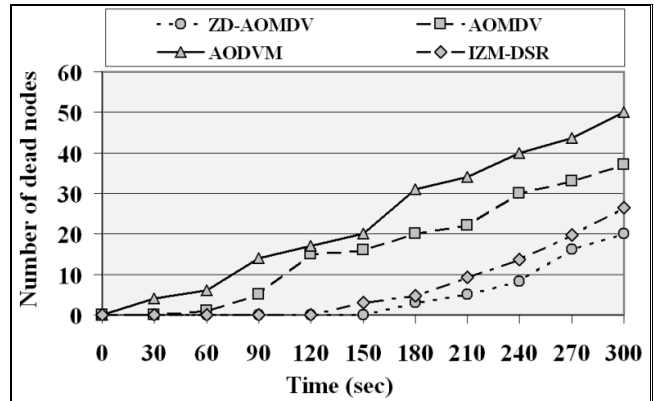

**Figure 21.** Number of Dead Nodes in Simulation Time.

*E) Energy Consumption*

In figures 22, 23 and 24 energy consumption are shown as a function of mobility speed and data rate. In ZD-AOMDV and IZM-DSR all discovered routes are used simultaneously thus many of nodes participate in forwarding data and their consumed energies are increased simultaneously, but in AODVM and AOMDV only one route is used for forwarding data thus only a few nodes are involved in transmitting data. In fact, ZD-AOMDV distributes energy consumption across many nodes thus the life time of network is increased but in AODVM and AOMDV the energy consumption is focused on a few nodes so the network life time is less than in ZD-AOMDV.

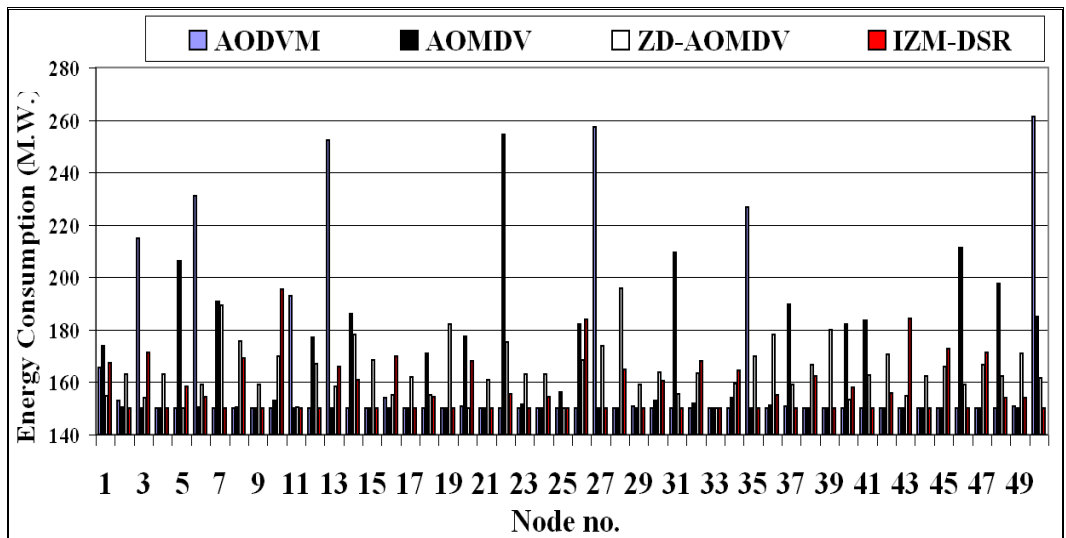

**Figure 22.** Energy Consumption
(Max speed=30Km/h, Pause time=15ms, No of Src=1, Terrafic=30Kbps)

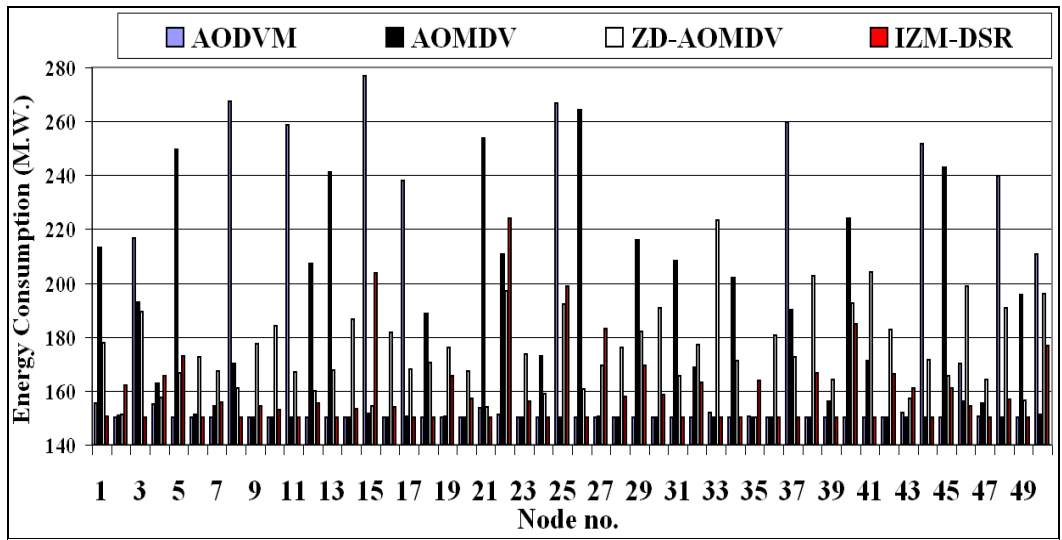

**Figure 23.** Energy Consumption
(Max speed=60Km/h, Pause time=30ms, No of Src=1, Terrafic=30Kbps)

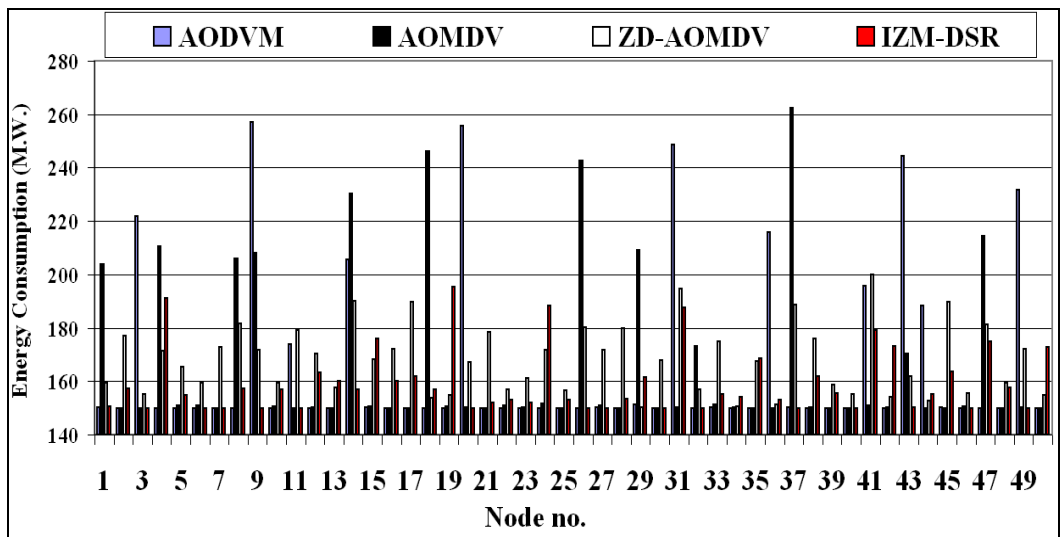

**Figure 24.** Energy Consumption
(Max speed=30Km/h, Pause time=15ms, No of Src=1, Terrafic=60Kbps)

With getting more thorough in figure 10, we can realize that the increase of the number of nodes has a very large effect on ZD-AOMDV overhead routing .This is because of the neighboring nodes increase and increase of the number of packets per query and query response in the path discovery process.

## 5. CONCLUSION

Some multipath algorithms in the ad hoc networks divide data at source and simultaneously send the different parts to destination via different paths to reduce end-to-end delay. In this way, using node disjoint paths seems like a good option. But sending traffic through node disjoint paths is not completely independent of each other and because of the mechanisms for shared channel access in wireless networks such as the CSMA/CA protocol, sending information through a path can affect another path. Such problems can be solved by implementing regional disjoint paths instead of node disjoint paths for sending information concurrently. In this paper, a new multipath routing algorithm is suggested based on AODV that uses all directional antenna to discover and use regional distinct paths. To achieve this goal, active neighbors of each path are counted. Also, selection is executed based on the number of active neighbors.

The proposed algorithm is compared to AOMDV, AODVM and IZM-DSR algorithms during various scenarios, and improvements are obtained in the field of energy consumption, end-to-end delay and packet delivery ratio. But instead, our proposed algorithm's routing overhead is higher than AOMDV and AODVM algorithms.